\documentclass[lettersize,journal]{IEEEtran}
\usepackage{amsmath,amsfonts}
\usepackage{algorithmic}
\usepackage{algorithm}
\usepackage{array}
\usepackage{textcomp}
\usepackage{stfloats}
\usepackage{url}
\usepackage{verbatim}
\usepackage{graphicx}
\usepackage[tight,footnotesize]{subfigure}
\usepackage{cite}
\usepackage{booktabs}

\usepackage{color} 

\usepackage{multicol} 
\begin{document}


\title{AMPLE: An Adaptive Multiple Path Loss Exponent Radio Propagation Model Considering Environmental Factors}
\author{Lingyou~Zhou,~\emph{Student Member, IEEE,} Jie~Zhang,~\emph{Senior Member, IEEE,} Jiliang~Zhang,~\emph{Senior Member, IEEE,} Oktay Cetinkaya,~\emph{Member, IEEE,} and Steve Jubb
	\vspace{-0.2in}
	
	\thanks{
		Lingyou Zhou is with the Department of Electronic and Electrical Engineering, the University of Sheffield, Sheffield, S10 2TN, UK.
		
		Jie Zhang is with the Department of Electronic and Electrical Engineering, the University of Sheffield, Sheffield, S10 2TN, UK, and also with Ranplan Wireless Network Design Ltd., Cambridge, CB23 3UY, UK. 
		
		Jiliang Zhang is with the College of Information Science and Engineering, Northeastern University, Shenyang, 110819, China.
		
		Oktay Cetinkaya is with the Department of Engineering Science, University of Oxford, Oxford, OX1 2JD, UK.
		
		Steve Jubb is with the Department of Civil and Structural Engineering, the University of Sheffield, Sheffield, S10 2TN, UK.
		
		
	}
}
\markboth{IEEE}%
{Shell \MakeLowercase{\textit{et al.}}: A Sample Article Using IEEEtran.cls for IEEE Journals}


\maketitle

\begin{abstract}
	
We present AMPLE -- a novel multiple path loss exponent (PLE) radio propagation model that can adapt to different environmental factors. The proposed model aims at accurately predicting path loss with low computational complexity considering environmental factors. In the proposed model, the scenario under consideration is classified into regions from a raster map, and each type of region is assigned with a PLE. The path loss is then computed based on a direct path between the transmitter (Tx) and receiver (Rx), which records the intersected regions and the weighted region path loss. To regress the model, the parameters, including PLEs, are extracted via measurement and the region map.  We also verify the model in a suburban area. To the best of our knowledge, this is the first time that a multi-slope model precisely maps PLEs and region types. Besides, this model can be integrated into map systems by creating a new path loss attribute for digital maps. 
	
	
\end{abstract}


\section{Introduction}

\IEEEPARstart{R}{adio} propagation loss characterization plays a crucial role in wireless network planning and optimization. Currently, deterministic and empirical models are the two most widely used categories of radio propagation models. Deterministic models, such as ray tracing and ray launching models \cite{he018}, compute rays from Rx to Tx and from Tx to Rx based on ray optics, respectively. To maintain high prediction accuracy, however, these models require a large amount of computation resources and time. Empirical models, such as Okumura-Hata model \cite{rap96}, use one characterization with a few parameters to calculate path loss for a typical scenario. Whereas, accuracy is sacrificed for maintaining minimum complexity and the type of scenarios is constrained to what the models are trained for. Moreover, for multi-slope models \cite{rap96, zha15}, even they divide the propagation into several distance-based parts corresponding to different PLEs, the precise PLE-region type mapping is missing.

Meanwhile, to combine the advantages of accuracy offered by deterministic models with the low computational complexity provided by empirical models, several semi-deterministic models for path loss prediction have been proposed \cite{md6}. The parameters in these models can be either extracted from measurement data, or supplied by ray-tracing simulation results. Among these models, the 3GPP channel model, the mmMAGIC channel model, the METIS channel model, the MiWEBA channel model, the 5GCMSIG and the IMT-2020 channel model accommodate propagation characteristics of blockage and gaseous absorption which can be extended to incorporate the high blockage effect due to environmental factors. However, an adaptive multiple PLE radio propagation model, which can automatically adapt its model parameters to different environmental factors, does not exist. 

In this paper, we present a fast Adaptive Multiple Path Loss Exponent (AMPLE) radio propagation model considering environmental factors. The AMPLE model first extracts the environment data from a raster map, by recognizing different environmental obstacles, the raster map is classified into multiple region types. To parametrize the environment, each region type is labeled with an undefined PLE that can be further determined via measurement. For path loss, we first assume that there is a direct path between Tx and Rx, and record all regions intersected with this path; then, we obtain the total path loss value in decibels by accumulating the path loss of each of the intersected regions with the direct path. We also apply the AMPLE model in a suburban area with simplified map and the results show the performance of the AMPLE model, with the root mean square error (RMSE) of 6.787 dB. To the best of our knowledge, this is the first time that the PLEs precisely corresponds to different region types in a multi-slope model. Also, this simple-but-accurate model can be integrated into map systems by creating a new path loss attribute for digital maps.

The rest of the paper is organized as follows. Section \ref{Sec2} describes the details of the AMPLE model. Section \ref{Sec3} demonstrates the AMPLE model in a suburban area. Section \ref{Sec4} summarizes and analyzes the parameter results of the AMPLE model in the measurement area. Section \ref{Sec6} presents conclusions and future works.

\section{The Model}
\label{Sec2}

\begin{figure*}[t]
	\centering
	\includegraphics[width = 0.85\linewidth]{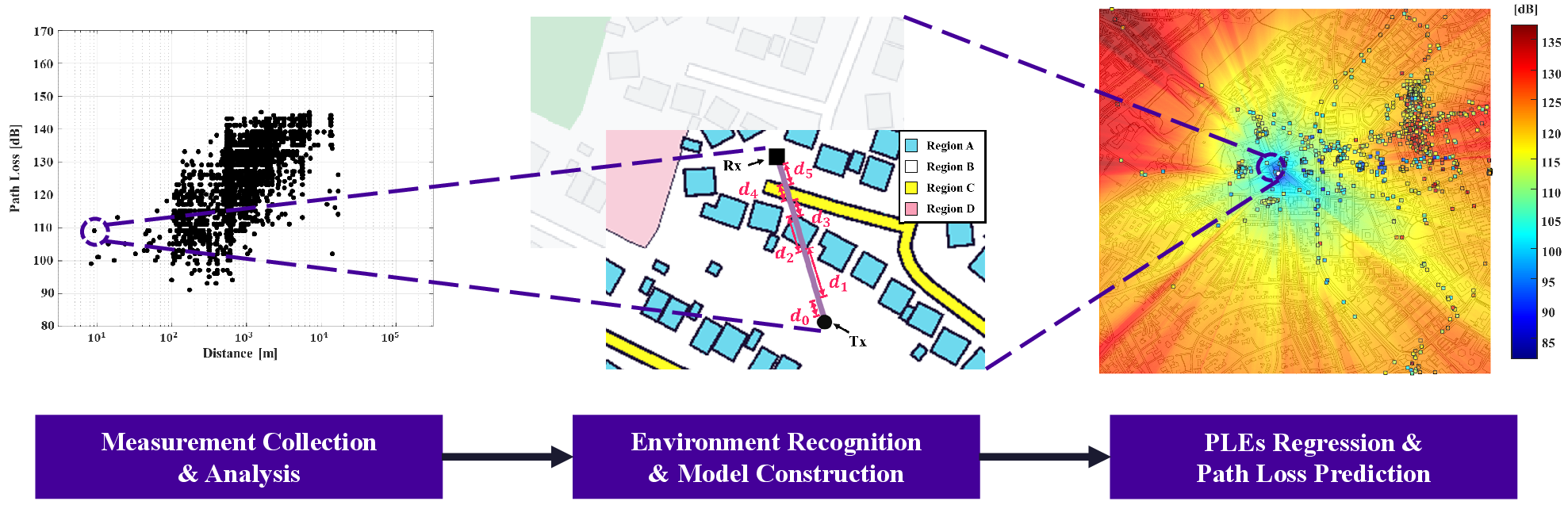}
	\caption{The process of constructing the AMPLE model.} 
	\label{PEFsPrcs}
	\vspace{-10pt}	
\end{figure*}
In this section, we first describe the AMPLE path loss model, wherein each PLE stands for the parameter of an environmental obstacle type (e.g., building, as a region type), and present the details of computing the environment-based path loss. Then, we reveal the relationship between the AMPLE model and the log-distance model under special conditions. We also give a discussion about the AMPLE model.

\subsection{Map Classification with Direct Path}
\begin{figure}[hb]
	\centering
	\includegraphics[width = 1\linewidth]{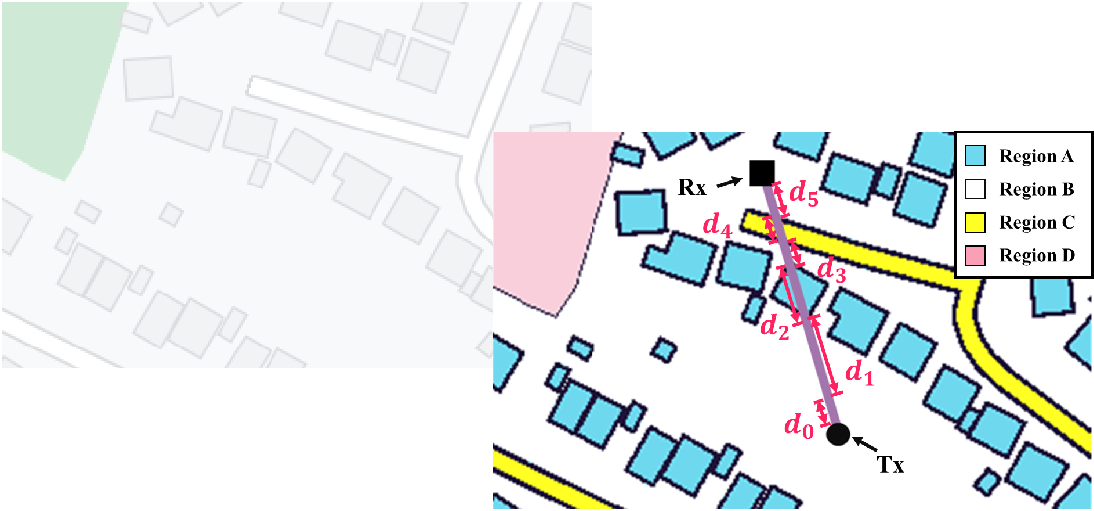}
	\caption{One example of raster map classification, from the original raster map to the region map, along with lengths of the intersected regions (e.g., $d_{0}$ and $d_{1}$ are the close-in distance and length of the first region, respectively).}
	\label{DsnExpl}		
\end{figure}

A typical modeling process of the AMPLE model is shown in Fig. \ref{PEFsPrcs}. Based on measurements, the path loss can be computed to serve as the foundation for model characterization. To incorporate environmental factors, a color-coded raster map of the measurement area is essential, which can be collected from map systems (e.g., Google Maps). Using k-means clustering based on the Euclidean distance of red, green and blue (RGB) values, environmental obstacles are classified into multiple region types, with an example shown in Fig. \ref{DsnExpl} (four types of regions are classified). Note that region types are assigned distinct PLEs to construct the path loss model. With the Tx-Rx direct path, the intersected regions and the corresponding region lengths are recorded, which is given by
\begin{equation}
	\label{LnMtx}
	\mathbf{S} =
	\begin{gathered}
		\begin{bmatrix}
			R_0  & R_1 & R_2    & R_3    & \cdots  & R_{N}\\
			d_0 & d_1 & d_2    & d_3    & \cdots  & d_{N} \\
		\end{bmatrix},
		\quad
	\end{gathered}
\end{equation}
where the first and second rows in \eqref{LnMtx} represent the intersected regions and the corresponding region lengths. Note that length $d_0$ is the close-in distance in the direct path corresponding to the close-in region $R_0$ \cite{mir13}, where regions within length $d_0$ are not counted. That is, the total number of regions intersected by the direct path is the numerical value $N$, rather than $N+1$.

Comprehensively, the direct path, starting from the close-in region $R_{0}$ and ending at the last region $R_{N}$, intersects $N$ regions in total (excluding the close-in region $R_0$). In Fig. \ref{DsnExpl}, four types of regions are recognized from the original raster map. In this case, the direct path intersects five regions (i.e., $N=5$), and the length of regions are shown from $d_0$ to $d_5$.

\subsection{The AMPLE Model}
The decibel characterization of the AMPLE model for each Tx-Rx direct path can be expressed as
\begin{equation}
	\label{PLPEFs}
	PL = C + \sum_{a=1}^{N} 10 n_a \log_{10}\left(\frac{\sum_{r=0}^{a} d_r}{\sum_{r=0}^{a-1} d_r}\right)+pX +\Psi_{\sigma},	
\end{equation} 
where $C$, $N$, $n_{a}$, $d_r$, $X$, $p$ and $\Psi_{\sigma}$ are characterized as follows.

\begin{itemize}	
\item{\textbf{\emph{Intercept}} ($C$): For the direct path, $C$ is the decibel path loss at close-in distance $d_0$  \cite{rap96,erc99}. Based on equation \eqref{PLPEFs}, regions within $d_0$ are not counted. Instead, $d_0$ is set as 1 m and the intercept $C$ is regressed along with PLEs.}

\item{\textbf{\emph{Intersected Regions}} ($N$, $n_a$): There are $N$ regions intersected by the Tx-Rx direct path, wherein $n_a$ is the corresponding PLE for $a$th region.} 

\item{\textbf{\emph{Weighted Distance}} ($d_r$): $d_r$ is the $r$th region length, and the fraction in parentheses is the weighted distance to compute the sub-path loss for $a$th region. The numerator and denominator are summations ranging from close-in distance $d_0$ up to the lengths of $a$th and $a$-1th region, respectively. Note that under the properties of logarithms, the fraction can be rewritten as a subtraction, which results in the sub-path loss of the desired region.}

\item{\textbf{\emph{Penetration Loss}} ($X$, $p$): $X$ is the penetration loss and $p$ is the total number of penetrations in the direct path. Two penetrations (i.e., in $\&$ out) are considered when the direct path intersects buildings, while the model remains unchanged for path passing through other region types (i.e., with corresponding PLE only). Note that penetrations are assumed to always occur for simplicity.}

\item{\textbf{\emph{Shadowing}} ($\Psi_\sigma$): The shadowing term $\Psi_{\sigma}$ is normally distributed in dB-scale with $N[0,{\sigma}^{2}]$.}
\end{itemize}
As the region types within the measurement area might not match the intersected regions $N$, equation \eqref{PLPEFs} can be simplified by combining terms with the same PLE. That is, if there are $M$ region types, the proposed model can be characterized as  
\begin{equation}
	\label{WtdPL_LkTrm}
	PL = C + \sum_{m=1}^{M} D_mn_m+pX+\Psi_{\sigma},
\end{equation} 
where $n_m$ is the $m$th region type PLE and $D_m$ is the corresponding coefficient. Also, as the model in dB-scale contains a shadowing term with normal distribution $N[0,{\sigma}^{2}]$, which means, the total path loss in this case is normally distributed with $N[\mu(C,n_{m},X),{\sigma}^{2}]$, where \cite{gus15}
\begin{equation}
	\label{MeanofWPL}
	\mu(C,n_{m},X) = C + \sum_{m=1}^{M} D_mn_m+pX.
	\vspace{-5pt}
\end{equation}

\subsection{Relationship between the AMPLE Model and the Log-distance Model}
From the above description, the AMPLE model is applicable to different scenarios according to different map-measurement sets. Thus, the quantity $M$ may increase for complex scenario (e.g., multiple region types) and decrease for simple scenario (e.g., single region type). Specially, for scenario with single region type or if we set $M$ to one, the characterization in \eqref{WtdPL_LkTrm} will be simplified to the log-distance model, which is
\begin{equation}
	\label{LogdPL}
	PL = C + 10n\log_{10}(d/d_{0})+\Psi_{\sigma},   \qquad     d \ge d_{0},
\end{equation}	
where $n$ is the PLE and $C$ is the intercept value that represents the decibel path loss at close-in distance $d_{0}$. For path loss prediction, this equation is common practice to characterize path loss in dB scale as the power-law of distance $n$ plus a shadowing term \cite{erc99}. Because the shadowing term is modeled by normal distribution in dB scale, the path loss in this model can be expressed as a normally distributed value with the same standard deviation, along with a distance-dependent mean, $N[\mu(d),{\sigma}^{2}]$, where \cite{gus15}
\begin{equation}
	\label{MeanofLgdPL}
	\mu(d) = C + 10n\log_{10}(d/d_{0}).
	\vspace{-5pt}
\end{equation}

\subsection{Discussion of the AMPLE Model}
For map classification, only a basic machine learning method is used (i.e., k-means clustering), since the map data has been pre-processed (e.g., Google Maps). For general cases (e.g., raw map data like satellite maps) with arbitrary region shapes, deep-learning (DL) methods such as convolutional neural networks (CNNs) can be considered as  alternatives \cite{min22}.
 
Although the propagation between Tx and Rx is depicted as a direct path shown in Fig. \ref{DsnExpl}, all the radio propagation mechanisms are incorporated by different PLEs. Along with the model simplicity in \eqref{PLPEFs} and \eqref{WtdPL_LkTrm}, the AMPLE model achieves fast but accurate path loss prediction. Without changing construction, the AMPLE model can cover different scenarios according to different map-measurement sets, and it can be further used in the stochastic channel models (e.g., the 3GPP channel model \cite{md6}) for path loss prediction.

To the best of the our knowledge, this is the first time that a multi-slope path loss model precisely maps PLEs to different region types. Beyond radio propagation, a new path loss attribute for digital map system is created by integrating the AMPLE model into map systems. The developments in digital maps and machine learning have made the adoption of such a model feasible.

\section{Model Application}
\label{Sec3}
In this section, we provide the application process of the AMPLE model in a suburban area with measurement. We first present the data collection method within the measurement area. Then, we describe the mapping between the region map and the data coordinates. In addition, the regression of model parameters based on truncated data is given.  

\vspace{-5pt}
\subsection{Measurement Collection}

\begin{figure}[t]
	\vspace{-1pt}	
	\centering
	\includegraphics[width = 0.75\linewidth]{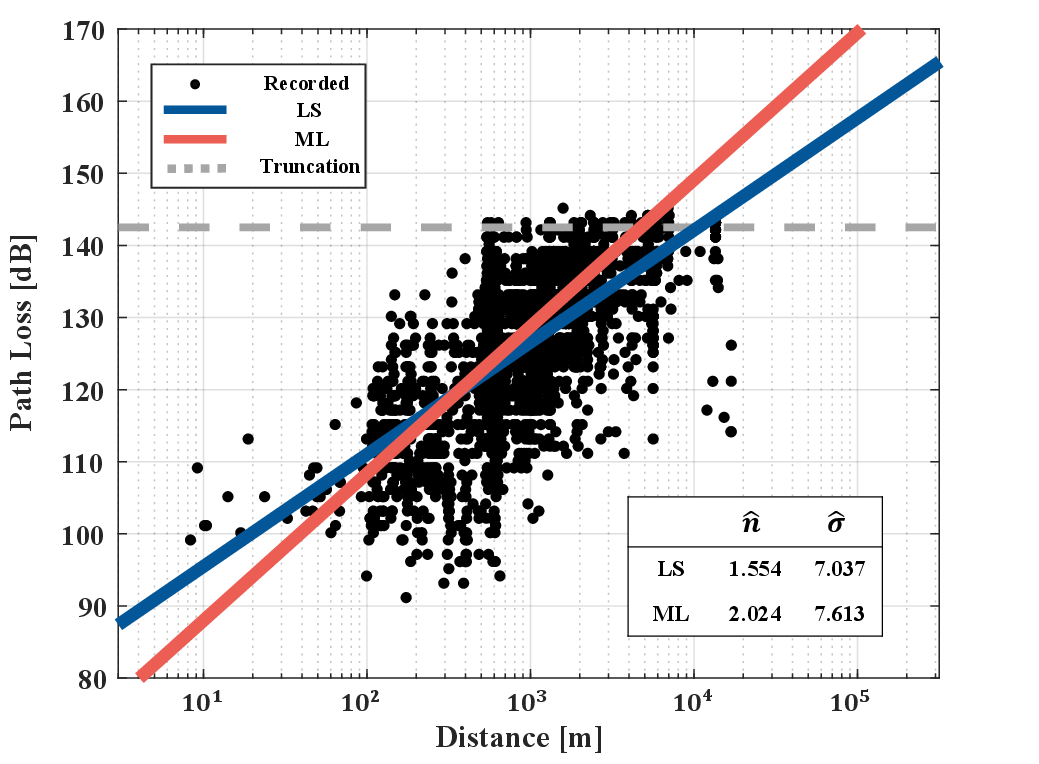}
	\vspace{-5pt}
	\caption{The scatter plot of measurement and estimation results of least-squares (LS) and maximum likelihood (ML) with truncated normal distribution.} 
	\vspace{-8pt}
	\label{SctPlt&CF}
\end{figure}
The measured data were collected in Sheffield, UK, considered as a suburban area with hilly terrain, dense buildings, and moderate-to-heavy tree densities \cite{erc99}. We focus on a gateway of Long Range Wide Area Network (LoRaWAN) \cite{lora19} that accommodates a LORIX One with a 4.150 dBi omnidirectional antenna. It is located on the Hicks Building (53.381029, -1.4864733) of University of Sheffield, which has a height of 30 m. For the mobile antenna, Sheffield City is mapped through  “drive test” experiments, which were performed with autonomous mobile network scanner and field test device (having an antenna gain of 2 dBi), installed on an electric vehicle (an average height of 1.2 m). The received signal strength indicator (RSSI) are periodically recorded, together with the global positioning system (GPS) coordinates. 

The measurement data collected in Sheffield City comprises 4615 data points transmitted around 868 MHz, with a transmitted power of 20 dBm. These data were taken at Tx-Rx separation distances ranging from several meters to 10 km. Since the antenna gains are included in RSSI \cite{lora19}, the path loss is calculated by subtracting the antenna gains of 6.150 dBi (4.150 dBi for Tx and 2 dBi for Rx) to exclude antenna effects.

Fig. \ref{SctPlt&CF} shows the scatter plot of 4615 measured path loss in meters. The gray dash line shows the truncation of measured data, while two straight lines represents two regression fit methods based on the log-distance model. Since the Rx detect limitation, path loss data beyond 142 dB is missing, leading to a bias in the estimated PLE and standard deviation of shadowing. This is evident in the blue line representing least squares (LS) in Fig. \ref{SctPlt&CF}, resulting in $\hat{n}=1.554$ and $\hat{\sigma}=7.037$. To mitigate this, maximum likelihood (ML) with truncated normal distribution is employed, yielding more accurate results: $\hat{n}=2.024$ and $\hat{\sigma}=7.613$ \cite{gus15, coh91}, as indicated by the red line in Fig. \ref{SctPlt&CF}. Based on equations \eqref{LogdPL} and \eqref{MeanofLgdPL}, the probability density function (PDF) of $z$th measured path loss is given by \cite{gus15, coh91}
\begin{equation}
		\label{PDFlogd}
		\begin{aligned}
			P(l_z;\mu(d),\sigma) =
			\left\{
			\begin{array}{ccc}
				\frac{1}{\sqrt{2\pi}\sigma}\frac{\mathrm{exp}\big(-\big(\frac{l_z-\mu(d)}{\sqrt{2}\sigma}\big)^{2}\big)}{\Phi\big(\frac{L-\mu(d)}{\sigma}\big)}  &{l_z<L}, \\
				0 & {\text{else}},
			\end{array} \right.
		\end{aligned}	
\end{equation}
where $L$ is the right truncated value as 140 dB, and $\Phi$ is the cumulative distribution function (CDF) of the standard normal distribution. 


\vspace{-6pt}
\subsection{Region Classification and Coordinates Mapping}
\begin{figure}[t]
	\centering
	\includegraphics[width = .9\linewidth]{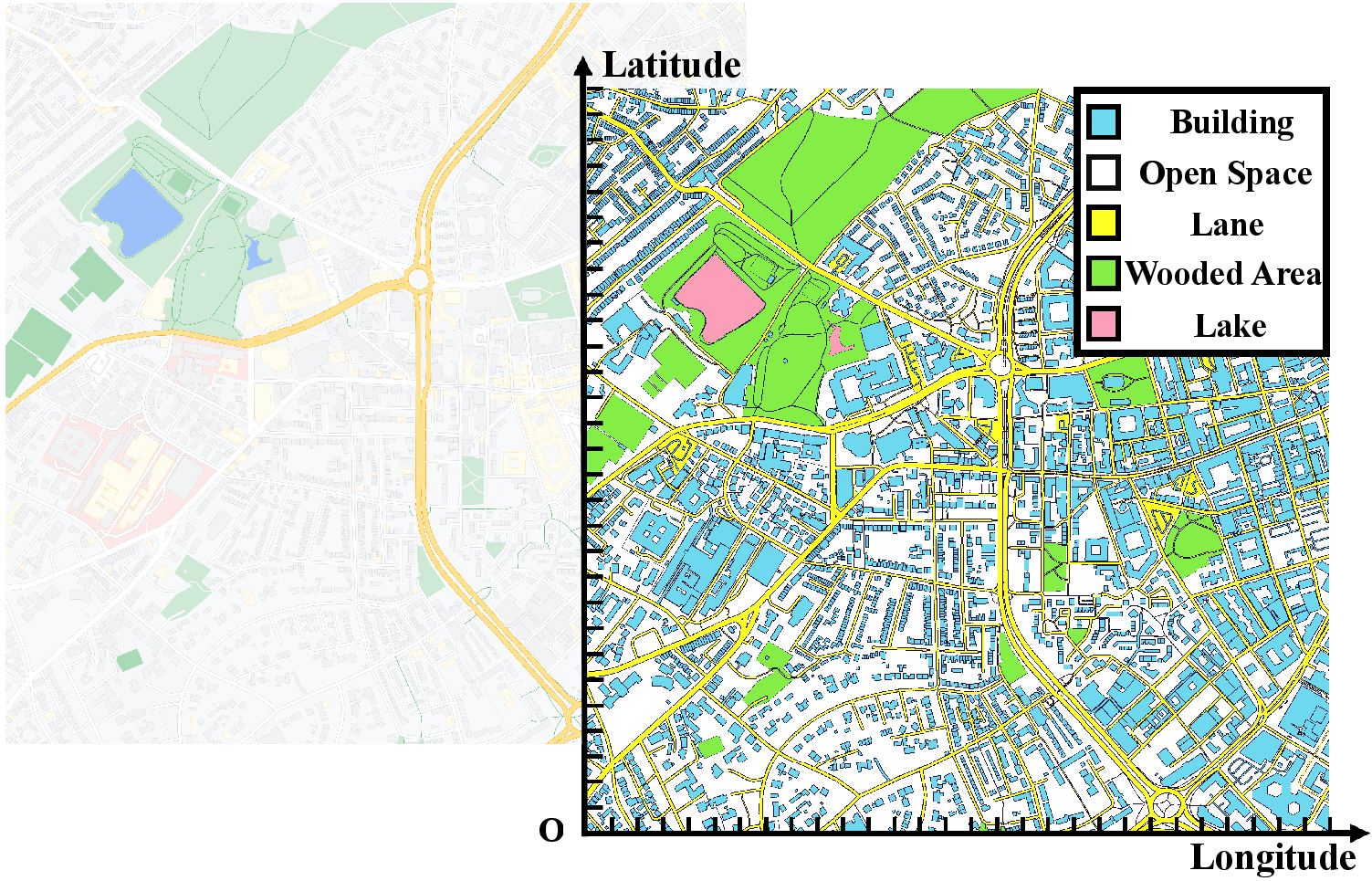}
	\caption{Map classification in partial Sheffield City, from the original raster map to the region map, along with the pixel coordinate system.}
	\label{figrastermapSfd}	
	\vspace{-10pt}
\end{figure}

The square raster map of partial Sheffield City is taken from Google maps, with a map scale of 50 m and the area of approximately 2.250 km$^{2}$ (has a side of 2115 pixels), as shown in Fig. \ref{figrastermapSfd}. Under the map constraints, 3128 out of 4615 measured path loss are included. The environmental obstacles are classified into the following region types:  $Building$, $Open\,Space$, $Lane$, $Wooded\,Area$ and $Lake$.  

Since the coordinates of Rx were recorded during experiments, the mapping between region map and measurement is to link coordinates with region map. Based on the coordinates of map edges, the coordinate system is established and illustrated in Fig. \ref{figrastermapSfd}, where each pixel can be constrained by its corner coordinates. For each Tx-Rx pair, the linear function of coordinates is generated to further record the intersected pixels. With exclusive-or algorithm, the intersected pixels can be finally constructed into intersected regions.

\vspace{-3pt}
\subsection{Parameter Regression with Truncated Data}
With region type $M$ equals five and equation \eqref{MeanofWPL}, we first calculate the PDF of \eqref{WtdPL_LkTrm} to process the truncation by ML with truncated normal distribution, which can be expressed as

\begin{footnotesize}
		\vspace{-5pt}
	\begin{equation}
		\label{PDF}
		\begin{aligned}
			P(l_z;\mu_{\mathrm{s}}(C,n_{m},X),\sigma) =
			\left\{
			\begin{array}{ccc}
				\frac{1}{\sqrt{2\pi}\sigma}\frac{\mathrm{exp}\big(-\big(\frac{l_z-\mu(C,n_{m},X)}{\sqrt{2}\sigma}\big)^{2}\big)}{\Phi\big(\frac{L-\mu(C,n_{m},X)}{\sigma}\big)}  &{l_z<L}, \\
				0 & {\text{else}},
			\end{array} \right.
		\end{aligned}
	\vspace{5pt}
	\end{equation}
\end{footnotesize}where the values in \eqref{PDF} are mentioned before. To estimate the quantities of \eqref{PDF} by ML, the partial derivatives of the likelihood function for all the values are required, including PLEs $n_m$, intercept $C$, penetration loss $X$, and the standard deviation $\sigma$, which are 

\begin{footnotesize}
\begin{equation}
	\label{PD_nm}
	\begin{aligned}
		\frac{\partial F(n_m,C,X,\sigma)}{\partial n_m} =&
		\sum_{z=1}^{Z} D_m \bigg[ \frac{\mu(C,n_{m},X)-l_z}{\sigma^{2}}\\ &-\frac{\mathrm{exp}\big(-\big(\frac{L-\mu(C,n_{m},X)}{\sqrt{2}\sigma}\big)^{2}\big)}{\sigma\int_{-\infty}^{ \frac{L-\mu(C,n_{m},X)}{\sigma}} \mathrm{exp}(-\frac{t^{2}}{2})dt} \bigg],
	\end{aligned}
\end{equation}  
\end{footnotesize}

\begin{footnotesize}
\begin{equation}
	\label{PD_C}
	\begin{aligned}
		\frac{\partial F(n_m,C,X,\sigma)}{\partial C}  =& \sum_{z=1}^{Z} \bigg[ \frac{\mu(C,n_{m},X)-l_z}{\sigma^{2}}\\ 
		&-\frac{\mathrm{exp}\big(-\big(\frac{L-\mu(C,n_{m},X)}{\sqrt{2}\sigma}\big)^{2}\big)}{\sigma\int_{-\infty}^{ \frac{L-\mu(C,n_{m},X)}{\sigma}} \mathrm{exp}(-\frac{t^{2}}{2})dt} \bigg],
	\end{aligned}
\end{equation} 
\end{footnotesize}

\begin{footnotesize}
\begin{equation}
	\label{PD_X}
	\begin{aligned}
		\frac{\partial F(n_m,C,X,\sigma)}{\partial X}  =& \sum_{z=1}^{Z} p\bigg[ \frac{\mu(C,n_{m},X)-l_z}{\sigma^{2}}\\ 
		&-\frac{\mathrm{exp}\big(-\big(\frac{L-\mu(C,n_{m},X)}{\sqrt{2}\sigma}\big)^{2}\big)}{\sigma\int_{-\infty }^{\frac{L-\mu(C,n_{m},X)}{\sigma}} \mathrm{exp}(-\frac{t^{2}}{2})dt} \bigg],
	\end{aligned}
\end{equation} 
\end{footnotesize}
and
\begin{footnotesize}
	\begin{equation}
		\label{PD_sig}
		\begin{aligned}
			\frac{\partial F(n_m,C,X,\sigma)}{\partial \sigma} &= \sum_{z=1}^{Z} \bigg[ \frac{1}{\sigma}-\frac{(\mu(C,n_{m},X)-l_z)^{2}}{\sigma^{3}} 
			\\ &- \frac{(L-\mu(C,n_{m},X)) \mathrm{exp}\big(-\big(\frac{L-\mu(C,n_{m},X)}{\sqrt{2}\sigma}\big)^{2}\big)}{ \sigma^{2} \int_{-\infty }^{\frac{L-\mu(C,n_{m},X)}{\sigma}} \mathrm{exp}(-\frac{t^{2}}{2})dt} \bigg], 
		\end{aligned}
	\end{equation}
\end{footnotesize}where $Z$ is the total data points (i.e., $Z=3128$). Based on \eqref{PD_nm}-\eqref{PD_sig}, the above parameters are combinatorially optimized by gradient descent with suitable step size.

\section{Results and Analysis}
\label{Sec4}

By setting $d_0$ to 1 m in \eqref{LnMtx} and applied the AMPLE model in the measurement area, the numerical values of PLEs are estimated combinatorially and shown in Table \ref{MulPLERst}. Also, the analysis of these numerical values are given as follows.
\begin{table}[ht]
	\caption{Model Parameters in the Measurement Area}
	\centering
	\begin{tabular}{ccc}
		\toprule
		\textbf{Environmental Factors}  & \textbf{Variables} & \textbf{Values} \\
		\midrule
		Intercept          & $C$ & 81.548  \\
		In-Building           & $n_1$ & 1.076 \\
		Open Space         & $n_2$ & 1.719\\
		Lane               & $n_3$ & 2.453 \\
		Wooded Area        & $n_4$ & 4.794 \\
		Lake               & $n_5$ & 1.081 \\
		Standard Deviation & $\sigma$ & 6.977\\
		Penetration Loss   & $X$ & 0.010 \\
		\bottomrule
	\end{tabular}
	\label{MulPLERst}
\end{table}
\begin{itemize}
	\item{\textbf{\emph{In-Building}}: Most of buildings in measurement area are teaching buildings and student accommodations (e.g., long corridors) with waveguide effect, which leads to much lower in-building PLE compare with a measured case that is 1.570 \cite{mir13}.}
	
	\item{\textbf{\emph{Open\,Space}}: For the measured base station, it is located at the top of Hicks Building, which is situated at the hillside. Combined with the waveguide effect caused by other buildings, the $n_2$ is lower than 2.}  
	
	\item{\textbf{\emph{Lane}}: For constantly moving vehicles (e.g., cars, buses and trams) on lanes which cause reflection, diffraction and scattering during the measurement \cite{rap96}, the value of $n_3$ is larger than 2.}
	
	\item{\textbf{\emph{Wooded\,Area}}: The intensive scattering caused by trees leads to severe loss, so the value of $n_4$ goes beyond 4.}
	
	\item{\textbf{\emph{Lake}}: When propagating through lakes at 868 MHz, the Sommerfeld-Zenneck surface waves are generated which is 10-20 dB stronger than space waves\cite{pet15}, \cite{kin85}. That is, the signal decays as the quantity of $n_5$ because it propagates over the lake surface as a circle instead of radiating through the air as a sphere.} 
\end{itemize}
\begin{figure}[t]
	\vspace{-5pt}
	\centering
	\subfigure[]{\includegraphics[width=.75\linewidth]{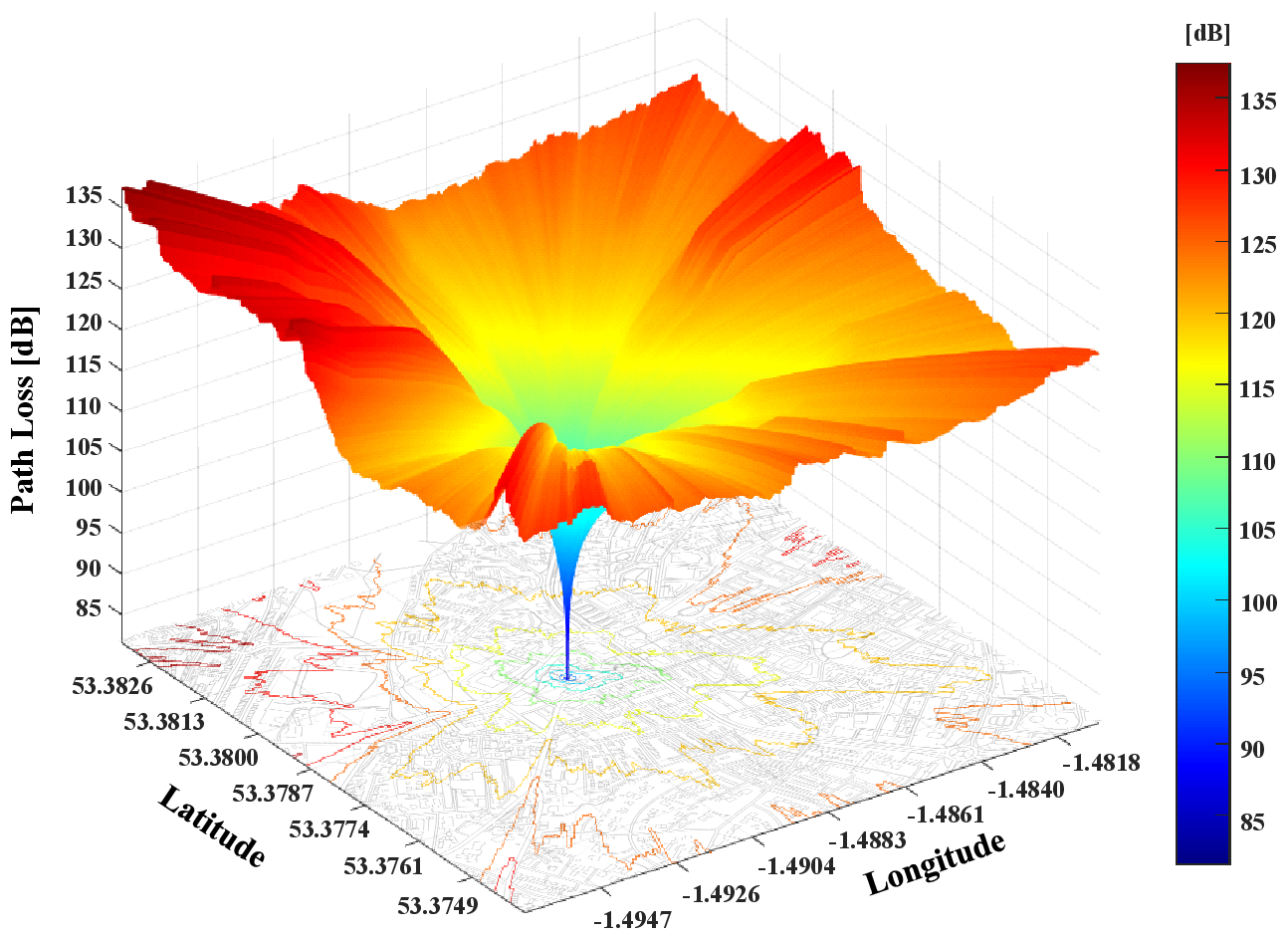}}%
	\hfil
	\subfigure[]{\includegraphics[width=.75\linewidth]{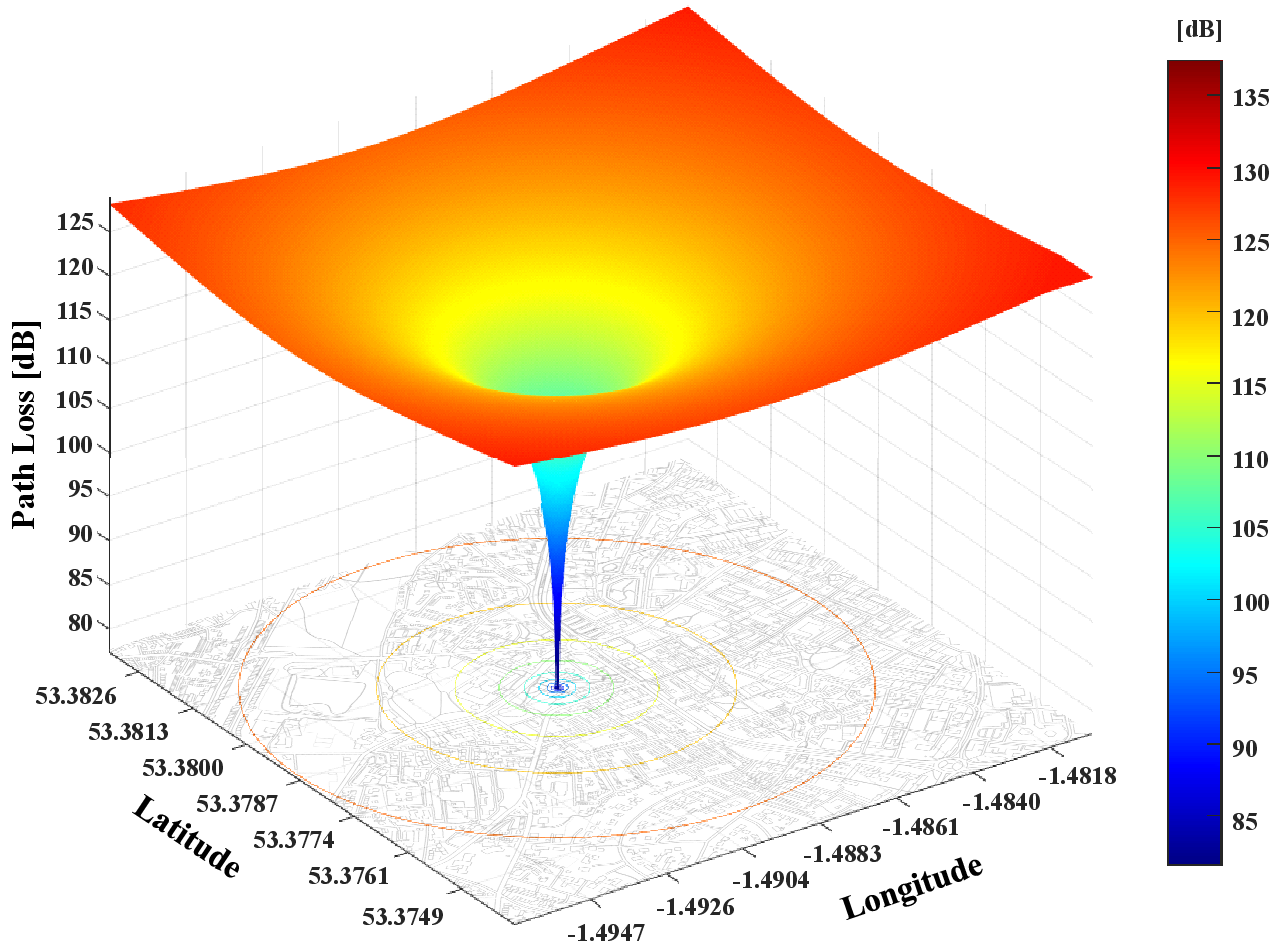}}%
	\caption{Heatmaps of the path loss prediction results predicted by (a) the AMPLE model and (b) the log-distance model.}
	\label{HeatMap}	
	\vspace{-15pt}
\end{figure}
Based on the values in Table \ref{MulPLERst}, we predict the path loss for the entire raster map by the AMPLE model, and compare the prediction results with the log-distance model. Note that the log-distance model is regressed by ML with truncated normal distribution in Section III, which can be expressed as 
\begin{equation}
	\label{LogdPLSheff}
	PL = 67.668+ 20.243\log_{10}(d)+\Psi_{\sigma},   \qquad     d \ge 1 \,\rm m,
\end{equation}	
where the standard deviation of shadowing term $\Psi_{\sigma}$ is $7.613$. In Fig. \ref{HeatMap}, the prediction of the AMPLE model and the log-distance model are shown in heatmaps, respectively. The measurement area is depicted with latitude and longitude axes, along with a bottom contour plot. As Tx-Rx distance increases, the path loss in the AMPLE model varies with intersected region types in Fig. \ref{HeatMap}(a), revealing regional influences towards path loss. Whereas, in Fig. \ref{HeatMap}(b), the contour plot of the log-distance model is shown as circles as distance increases. 

For the RMSE in dB scale, it is 7.098 dB in the log-distance model and 6.787 dB in the AMPLE model. The close proximity of RMSE values can be attributed to two reasons. First, map area constraints the variability of path loss predictions. As the Tx-Rx distance increases, variations between links become more pronounced. That is, region-based prediction surpass distance-based fast prediction. Second, the pre-processed maps only offer general geographical data, without the capability to differentiate sub-region types (e.g., $Wooded\,Area$ includes grassland and forest, which is unclassifiable by Google Maps).
 
\section{Conclusion and Future Works}
 \label{Sec6}
We have presented the AMPLE path loss model with multiple PLEs considering environmental factors. We also applied the AMPLE model in a suburban area and compare it with the log-distance model. The AMPLE model can not only be used to obtain fast and accurate path loss predictions, but can also be integrated into map systems by creating a new path loss attribute for digital maps. Therefore, the AMPLE model has the potential for wide applications and will have a significant impact in wireless communications.

For future works, the first is to reveal the factor of antenna heights. Since the model is based on two-dimensional maps, altering antenna heights requires more than PLE adjustments. The second is the factor of frequencies. As the major mechanisms of radio propagation may differ in different frequency bands, huge prediction errors may occur when deploying a path loss model to another frequency band. The third is the region classification in general cases. For raw map data with arbitrary region shapes, DL-based methods offer more precise map classification, yielding more detailed region types.

\vfill

\end{document}